# Atomic layer deposition of Y$_2$O$_3$ on *h*-BN for a gate stack in graphene FETs

[1]N. Takahashi, [2]K. Watanabe, [2]T. Taniguchi, and [1,3]*K. Nagashio

[1]*Department of Materials Engineering, The University of Tokyo,*
*7-3-1 Hongo, Bunkyo, Tokyo 113-8656 Japan*
[2]*National institute of Materials Science,*
*1-1 Namiki, Tsukuba, Ibaraki, 305-0044, Japan*
[3]*PRESTO, Japan Science and Technology Agency,*
*7-3-1 Hongo, Bunkyo, Tokyo 113-8656 Japan*

Corresponding authors:
*E-mail: nagashio@material.t.u-tokyo.ac.jp (K. Nagashio)

**Abstract:** The combination of *h*-BN and high-*k* dielectrics is required for a top gate insulator in miniaturized graphene field-effect transistors because of the low dielectric constant of *h*-BN. We investigated the deposition of Y$_2$O$_3$ on *h*-BN using atomic layer deposition. The deposition of Y$_2$O$_3$ on *h*-BN was confirmed without any buffer layer. An increase in the deposition temperature reduced the surface coverage. The deposition mechanism could be explained by the competition between the desorption and adsorption of the Y precursor on *h*-BN due to the polarization. Although a full surface coverage was difficult to achieve, the use of an oxidized metal seeding layer on *h*-BN resulted in a full surface coverage.

## 1. Introduction

The formation of an electrically reliable top-gate insulator on graphene is a key component of graphene field effect transistors (FETs) [1]. Recently, we demonstrated the high insulating properties of a Y$_2$O$_3$ top-gate in a graphene FET by applying high-pressure O$_2$ post-deposition annealing [2,3]. However, the direct deposition of high-*k* oxides generally degrades the transport properties of graphene. Alternatively, hexagonal boron nitride (*h*-BN) [4], a 2-dimensional crystal similar to graphite, has been considered to be an ideal insulator for extracting the intrinsic physical properties of graphene [5-7]. *h*-BN forms an atomically clear interface with graphene due to very weak van der Waals interactions. For the realistic integration of *h*-BN in transistor applications, the combination of *h*-BN and high-*k* dielectrics is required to reduce the effective oxide thickness (EOT) because the dielectric constant of *h*-BN is small (*k* is approximately 4). The key issue in this combination is to control the surface roughness for very thin high-*k* dielectrics.

One of the widely used deposition methods for gate dielectrics in graphene FETs is atomic layer deposition (ALD) because of the excellent conformality, the film-thickness controllability and the limited damage to graphene [8-10]. However, a nucleation layer is required for the ALD of dielectrics on graphene due to the chemical inertness of graphene surface [11-13]. This prevents from the potentially-high thickness control. The direct deposition of dielectrics on graphene without introducing defects to graphene still remains a challenge [14,15]. In contrast, it has been reported that Al$_2$O$_3$ can be directly deposited on *h*-BN and MoS$_2$ by ALD due to the in-plane electrical polarization of *h*-BN [16]. However, contradicting results have shown the non-uniform ALD of MoS$_2$ [17-19]. It is not clear whether the direct deposition of high-*k* dielectrics on *h*-BN is possible. In this study, Y$_2$O$_3$ was used as a high-*k* dielectric because of the lattice matching with graphene [20]. We report the deposition of Y$_2$O$_3$ on *h*-BN by ALD and discuss the deposition mechanism.

## 2. Experimental procedure

In this study, (i-PrCp)$_3$Y was used as a precursor [21-26]. Because the vapor pressure of (i-PrCp)$_3$Y is only 1/100 of trimethyl aluminum (TMA) at the room temperature, the container for (i-PrCp)$_3$Y was preheated to 130°C in a PICOSUN ALD reactor and the carrier gas flow rate in the chamber was increased to ensure a uniform film thickness for a 4 inch wafer. In this study, N$_2$ and water were used as a purge gas and an oxidant, respectively. To determine the optimum deposition conditions, Y$_2$O$_3$ was deposited on a p-Si wafer with a 5 nm-thick SiO$_2$ layer by ALD. The film thickness, surface roughness and amount of carbon impurities were characterized using spectroscopic ellipsometry, atomic force microscopy (AFM) and X-ray photoelectron spectroscopy (XPS), respectively. Moreover, the dielectric constant was estimated using capacitance-voltage (*C-V*) measurements. *h*-BN was mechanically exfoliated and transferred to a 90 nm-thick SiO$_2$/Si wafer. Because the chemical cleaning process for the tape residue often results in unintentional *h*-BN surface contamination, Y$_2$O$_3$ was deposited on *h*-BN at different temperatures from 25 °C to 400 °C without any cleaning processes. Alternatively, highly oriented pyrolytic graphite (HOPG) was used as a substrate for comparison.

## 3. Results & Discussion

First, the self-limiting surface growth conditions for Y$_2$O$_3$ were explored. **Figure 1(a)** shows the growth rate of Y$_2$O$_3$ as a function of precursor pulse time ($t_Y$) at a reactor temperature of 200°C. The time periods for the N$_2$ purge, water pulse and subsequent N$_2$ purge were fixed as 45 s, 0.1 s and 60 s, respectively. As the precursor pulse time increased, the growth rate initially increased and saturated at approximately 0.07 nm/cycle, which confirmed the self-limiting nature of ALD. It is noted that the growth rate increases again due to the decomposition of precursor with water in the reactor when the precursor pulse time increased further. Then, the precursor pulse time was fixed at 3 s, and the growth temperature was changed. The so-called constant growth rate ALD-window [27] was confirmed in **Fig. 1(b)**. Although it was not so wide, this is not so rare for some kinds of Y-precursors [23]. Under this optimized condition, the linear relation between the Y$_2$O$_3$ thickness and the number of ALD cycles was clearly observed, as shown in the inset of **Fig. 1(b)**. The growth rate was approximately 0.07 nm/cycle, which was consistent with previous studies using different precursors [24]. The root mean square (RMS) for the Y$_2$O$_3$ surface thickness was approximately 0.2 nm, which was consistent with that estimated for the initial SiO$_2$/Si surface.

Next, the dielectric constant of Y$_2$O$_3$ deposited by ALD was characterized. The *C-V* curves of Au/Y$_2$O$_3$/SiO$_2$/p-Si/Al capacitors were obtained at different frequencies from 1 kHz to 1 MHz. The inset in **Fig. 2** shows the typical *C-V* curves for the capacitor. No frequency dispersions were observed in the saturation region. **Figure 2** shows the inverse of the measured capacitances for the accumulation region at 1 MHz as a function of the $Y_2O_3$ thickness. The blue solid boxes indicate the mean values for 5 measured points. In this plot, the following equation can be applied,

$$\frac{1}{C_{meas}} = \frac{1}{C_{Y2O3}} + \frac{1}{C_{SiO2}} + \frac{1}{C_{Si}}$$
$$= \frac{1}{k_{Y2O3}\varepsilon_0} d_{Y2O3} + constant, \quad (1)$$

where $C_{meas}$, $C_{Y2O3}$, $C_{SiO2}$, $C_{Si}$, $\varepsilon_0$, and $d$ are the measured capacitance, $Y_2O_3$ capacitance, $SiO_2$ capacitance, Si capacitance, dielectric constant of vacuum, and thickness of $Y_2O_3$, respectively. Therefore, the $k$-value for $Y_2O_3$ was estimated to be 12.9 from the slope because it represents $1/(k_{Y2O3}\varepsilon_0)$. This $k$-value was nearly the same as that of bulk $Y_2O_3$ [28], which suggested the high electrical quality of $Y_2O_3$.

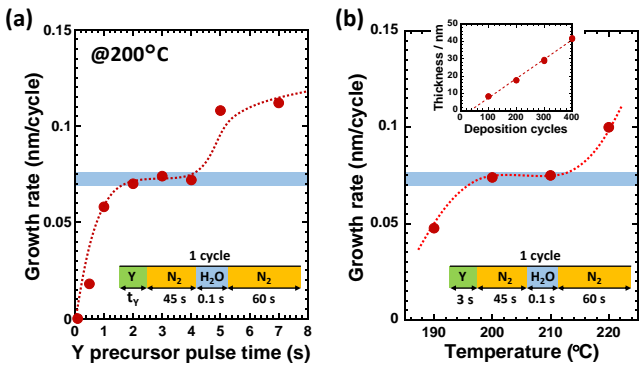

**Fig. 1** Growth rate of Y$_2$O$_3$ deposited on SiO$_2$/Si at 200 °C (a) as a function of Y precursor pulse time and (b) as a function of the deposition temperature. Y pulse time, N$_2$ purge time, H$_2$O pulse time, and N$_2$ purge time are indicated. The inset in (b) shows the Y$_2$O$_3$ thickness as a function of the deposition cycle at 200 °C.

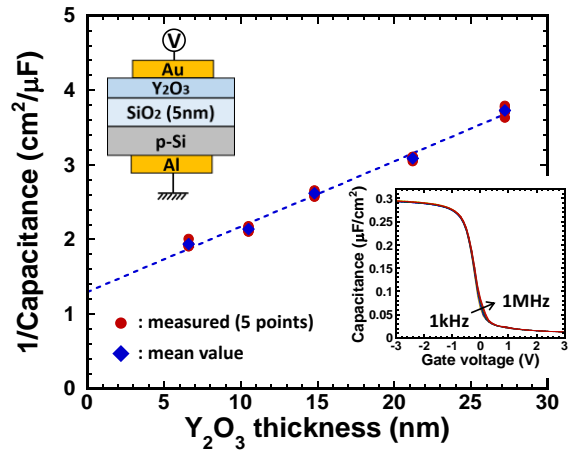

**Fig. 2** Inverse of measured capacitance for Au/Y$_2$O$_3$/SiO$_2$/p-Si/Al as a function of Y$_2$O$_3$ thickness. The right inset shows the typical *C-V* curves for the capacitor as a function of gate voltage at different frequencies from 1 kHz to 1 MHz.

In general, carbon impurity is one of main problems for ALD because the ligands of the precursors are mostly composed of carbon atoms. Moreover, it is known that $Y_2O_3$ can include OH group impurities [24]. The annealing after ALD is essential to remove both carbon and OH group impurities. Here, XPS data were collected for $Y_2O_3$ deposited on $SiO_2$/Si without annealing. **Figure 3** shows the peaks for (a) $C_{1s}$ and (b) $O_{1s}$ for $Y_2O_3$ with the thickness of ~7 nm deposited in the ALD mode (red curves), where the deposition temperature was 200 °C and the precursor pulse time was 3 s, as shown in **Fig. 1(a)**. The small peak of $C_{1s}$ can be detected at 284.9 eV, which is close to the binding energy for C-C. The carbon concentration is ~ 8 % from the area ratio. In **Fig. 3(b)**, the slightly larger peak of the OH group and the $O_{1s}$ peak for Y-O can be detected at 529.7 eV and 531.8 eV, respectively, which are consistent with the observations of a previous report [22,24]. It is noted that the contribution to $O_{1s}$ from Si-O can be neglected since the XPS signal for Si (103.9 eV) was negligible. Here, the $O_2$ annealing of $Y_2O_3$ on graphene at a high temperature (e.x. ≥ 350 °C) generally induces defects in graphene [2,3]. Therefore, to reduce the carbon and OH group impurities without annealing at the elevated temperature, the deposition condition was intentionally deviated from the ALD mode to the near CVD mode, where the decomposition of Y precursor in the reactor is facilitated. This could also be effective solution to the issue of steric hindrance on the substrate surface due to the large size of the Y precursor, because the Y precursor decomposes in the chamber. Experimentally, the precursor pulse time was extended from 3 to 5 s and the $N_2$ purge time after water pulse was shortened from 60 to 45 s. In this case (blue curves in **Fig. 3**), the $C_{1s}$ peak was negligible, which indicated that carbon impurities were undetectable by XPS. The OH group impurity was also considerably reduced and was at level similar to that after annealing at 400 °C [24]. Since the $k$ value of $Y_2O_3$ for near CVD mode was almost equal to that for ALD mode, the $Y_2O_3$ film quality for near CVD mode is better than that for ALD mode.

As mentioned above, $Y_2O_3$ was successfully deposited on $SiO_2$/Si. Next, we focused on the $Y_2O_3$ deposition on HOPG and $h$-BN. **Figures 4(a)** and **(b)** show the AFM images of $Y_2O_3$ on the HOPG and $h$-BN, respectively, where $Y_2O_3$ was deposited over 100 cycles at 150 °C. The optimum deposition temperature of $Y_2O_3$ on HOPG and $h$-BN was 150 °C, which is different from that on $SiO_2$. The detailed experimental data is shown later. A striped $Y_2O_3$ pattern was observed along the grain boundaries and defects on the HOPG due to the chemical inertness of the surface [11-13], whereas $Y_2O_3$ covered approximately 70% of the $h$-BN surface without any preferences for the grain boundaries and defects. These differences suggested that the deposition mechanisms were considerably different.

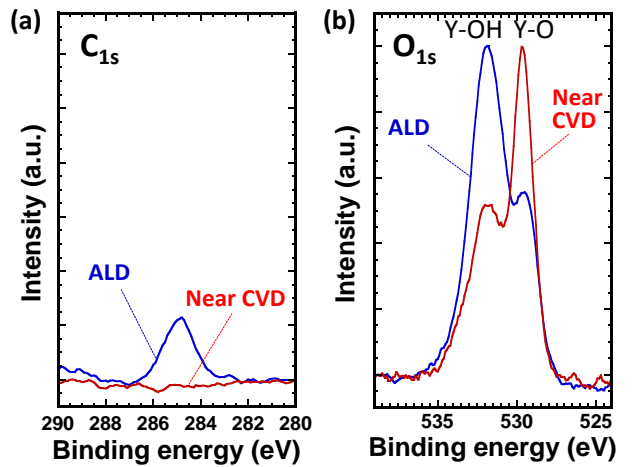

**Fig. 3** XPS spectra of (a) $C_{1s}$ and (b) $O_{1s}$ for $Y_2O_3$ deposited on $SiO_2$/Si at 200 °C. For $O_{1s}$, the intensities for both spectra are normalized due to different samples.

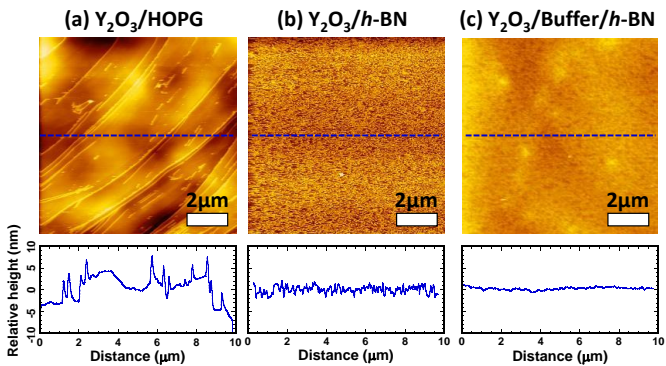

**Fig. 4** AFM images for (a) $Y_2O_3$ on HOPG, (b) $Y_2O_3$ on *h*-BN, and (c) $Y_2O_3$ on *h*-BN with oxidized Y metal buffer layer of 1.5 nm. Bottom figures show the relative height profiles along the dotted lines in AFM images.

To elucidate the deposition mechanism of $Y_2O_3$ on *h*-BN, $Y_2O_3$ and $Al_2O_3$ were deposited over 100 cycles at various temperatures, as shown in **Fig. 5(a)**. At 150°C, the deposition of $Y_2O_3$ was fairly uniform, but became more web-like with increasing temperature. Compared with $Al_2O_3$ on *h*-BN, $Y_2O_3$ disappeared at higher temperatures (300°C). Based on this temperature dependence, the deposition of $Y_2O_3$ on *h*-BN could be explained by the physical adsorption of the Y precursor on *h*-BN due to the polarization in *h*-BN because the physical adsorption of water on *h*-BN was reported to be small [16]. The desorption likely occurred at a high temperature because the thermal energy exceeded the activation barrier for the adsorption.

To compare the adsorption energies of two different precursors, we consider the molecular electronegativity ($S_m$) [29], which is expressed by

$$S_m = \left(\prod_{i=1}^{N} X_i\right)^{1/N}, \qquad (2)$$

where X is the electronegativity of the component atom. That is, the molecule electronegativity is the geometric mean of the individual electronegativities of all atom species of the molecule. The Pauling values for elemental electronegativity [30] and molecular electronegativity as estimated using eq. (2) are shown in **Table 1**. When the differences in the molecular electronegativities between precursors and *h*-BN were compared, (i-PrCp)$_3$Y/*h*-BN (0.18) was smaller than TMA/*h*-BN (0.27). Therefore, it was expected that the adsorption energy for (i-PrCp)$_3$Y on *h*-BN was smaller than that for TMA on *h*-BN. This was consistent with the higher surface coverage at high temperatures for $Al_2O_3$, as shown in **Fig. 5(a)**.

**Figure 5(b)** schematically summarizes this deposition mechanism. In this discussion, it was assumed that the adsorption energies for precursors on *h*-BN were fixed with temperature for simplicity. However, the desorption energy as the thermal energy generally increased with increasing temperatures. Moreover, the chemical decomposition of the precursor did not occur at a low temperature, as shown in **Fig. 5(a)**; this indicated a low temperature limit for ALD. Therefore, the surface coverage curve for the deposition can be expressed in **Fig. 5(b)**, based on the competition between the adsorption and desorption of precursors on *h*-BN.

Finally, we consider the potential for the full surface coverage of $Y_2O_3$ on *h*-BN by the adsorption mechanism. Although the full coverage of $Al_2O_3$ on *h*-BN has been reported by ALD without buffer layer [16], the direct deposition was not achieved in this study. In the previous study [16], *h*-BN was soaked in acetone, methanol and isopropanol to remove the tape residue. This cleaning process might leave the organic residue on *h*-BN, which act as nucleation sites [17]. Here, as discussed above, a difference existed in the molecular electronegativity between (i-PrCp)$_3$Y and *h*-BN.

Table I  Element electronegativity and molecule electronegativity

| Element | Y | Al | B | N | O | C | H |
|---|---|---|---|---|---|---|---|
| Electronegativity | 1.22 | 1.61 | 2.04 | 3.04 | 3.44 | 2.55 | 2.20 |
| Molecule | (i-PrCp)$_3$Y | | TMA | | *h*-BN | | |
| Molecule electronegativity | 2.31 | | 2.22 | | 2.49 | | |

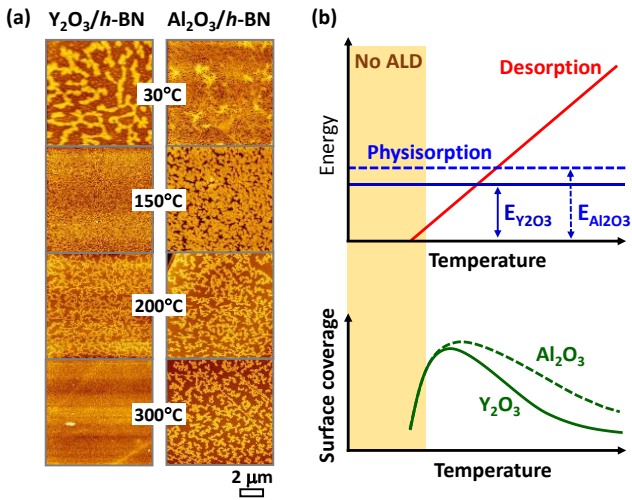

**Fig. 5** (a) AFM images for $Y_2O_3$ and $Al_2O_3$ deposited on *h*-BN at different temperatures. (b) Schematic diagram for the deposition mechanism of precursors. The surface coverage can be explained as the competition between adsorption and desorption at different temperatures.

However, the difference was not large, especially when considering that the atomic electronegativity for nonpolar C-H is 0.51. The thickness of $Y_2O_3$ deposited on *h*-BN decreased with increasing the temperature from 150 °C to 300 °C, while the thickness of $Y_2O_3$ deposited on $SiO_2$ increased with increasing the temperature. This suggested that desorption of Y precursor from the *h*-BN surface is considerably large. Moreover, the sizes of the precursors for the rare earth elements are generally much larger than that for TMA [27]; thus, the steric hindrance due to the large size of the Y precursor should be taken into account. Therefore, based on the above discussion, the full surface coverage by the direct deposition of $Y_2O_3$ on *h*-BN by ALD remains a challenge.

Here, to achieve the full surface coverage of $Y_2O_3$, the Y metal buffer layer of 1.5 nm was deposited on *h*-BN at 0.1 nm/s by thermal evaporation of Y metal in BN crucible under the Ar atmosphere of $10^{-1}$ Pa. Then, it was annealed in $O_2$ atmosphere at 200 °C for 10 min. Subsequent ALD deposition of $Y_2O_3$ resulted in a full surface coverage with a small surface roughness of approximately 0.2 nm, as shown in **Fig. 4(c)**.

## 4. Summary

We demonstrated the deposition of $Y_2O_3$ with $k = 12.9$ on $SiO_2$/Si wafer by exploring the deposition conditions. $Y_2O_3$ was successfully deposited directly on *h*-BN due to the physical adsorption of the Y precursor from the polarization in *h*-BN, unlike on HOPG. The surface coverage mechanism can be explained by the competition between adsorption and desorption at different temperatures. Although a full surface coverage is difficult to achieve due to the small adsorption energy and the steric hindrance of the large-sized Y precursor, an oxidized metal seeding layer on *h*-BN can provide for the full surface coverage of $Y_2O_3$. These results will provide an important step to the hetero-multi-layered gate-stacking of $Y_2O_3$/h-BN/graphene in graphene FETs.

## Acknowledgements

The authors acknowledge Prof. Toriumi and Dr. Nabatame for their helpful suggestions. This research was partly supported by a Grant-in-Aid for Scientific Research on Innovative Areas, for Young scientists, and for Challenging Exploratory Research by the Ministry of Education, Culture, Sports, Science and Technology in Japan.